\documentclass[reprint, prb]{revtex4-1}
\pdfoutput=1
\usepackage[bookmarks=false,colorlinks,citecolor=red]{hyperref}

\usepackage{graphicx}
\usepackage{color}
\usepackage{dcolumn} % Align table columns on decimal point
\usepackage{bm}      % bold math
\usepackage{amssymb}
\usepackage{amsfonts}
\usepackage{amsmath}

\usepackage{caption}
\usepackage{tikz} %circle numbers
\include{MyCommand}

\newcommand{\COMMENTED}[1]{}

\begin{document}

\title{Auxiliary Field Quantum Monte Carlo for Multiband Hubbard Models: Controlling the Sign and Phase Problems to Capture Hund's Physics}

\author{Hongxia Hao}
\affiliation{Department of Chemistry, Brown University, Providence, RI 02912}
\author{Brenda M. Rubenstein}
\affiliation{Department of Chemistry, Brown University, Providence, RI 02912}
\author{Hao Shi}
\affiliation{Center for Computational Quantum Physics, The Flatiron Institute, 162 5th Avenue, New York, NY 10003}

\begin{abstract}
In the study of strongly-correlated, many-electron systems, the Hubbard Kanamori (HK) model has emerged as one of the prototypes for transition metal oxide physics. The model is multi-band in nature and contains Hund’s coupling terms, which have pronounced effects on metal-insulator transitions, high temperature superconductivity, and other physical properties. In the following, we present a complete theoretical framework for treating the HK model using the ground state Auxiliary Field Quantum Monte Carlo (AFQMC) method and analyze its performance on few-band models whose parameters approximate those observed in ruthenate, rhodates, and other materials exhibiting Hund's physics. Unlike previous studies, the constrained path and phaseless approximations are used to respectively control the sign and phase problems, which enables high accuracy modeling of the HK model's ground state properties within parameter regimes of experimental interest. We demonstrate that, after careful consideration of the Hubbard-Stratonovich transformations and trial wave functions employed, relative errors in the energy of less than $1\%$ can routinely be achieved for moderate to large values of the Hund's coupling constant. Crucially, our methodology also accurately predicts magnetic ordering and phase transitions. The results presented open the door to more predictive modeling of Hund's physics within a wide range of strongly-correlated materials using AFQMC.  
\end{abstract}

\maketitle

% --------------------------------------------------------------------------------------
\section{Introduction}

%\HS{Reminder to Hao Shi: check citation careful, add more.}
Since the unanticipated discovery of high-temperature superconductivity in the cuprates, the single-band Hubbard model \cite{Hubbard_PRSLA_1993} has been the focus of an unparalleled level of theoretical scrutiny and associated algorithmic development.\cite{LeBlanc_Gull_PRX_2015,Lieb_Wu_RPL_1968,Scalapino_Brooks_NY_2007,Gukelberger_Werner_PRB_2015} Nevertheless, most materials exhibiting strong correlation, including most transition metal oxides\cite{KuneA_Pickett_NM_2008,Laad_Hartmann_PRB_2006,Maeno_Ikeda_MSE_1999} as well as the pnictides,\cite{Si_Abrahams_NRM_2016,Yin_Kotliar_NM_2011,Haule_Kotliar_NP_2009} fullerides,\cite{Nomura_JPCM_2016,Nomura_Science_2015} and chalcogenides\cite{Sun_Zhao_Nature_2012,Si_Abrahams_NRM_2016,Yin_Kotliar_NM_2011} possess multiple bands that cross their Fermi levels and are therefore fundamentally multi-band in nature.\cite{Georges_AnnRev_2013} In recent years, it has become increasingly evident that some of the most significant effects in such multi-band materials stem from Hund's coupling.\cite{Yin_Kotliar_NM_2011,Haule_Kotliar_NP_2009,Johannes_Mazin_PRB_2009} According to Hund's rules, electrons favor maximizing their total spin by first occupying different, degenerate bands in the same shell with parallel spins; only after they fill all available bands do they then doubly occupy the same bands.\cite{Georges_AnnRev_2013} As such, the effective Coulomb repulsion among electrons in a half-filled shell is increased due to Hund's rules, while that at any other filling is decreased. Hund's effects therefore drive half-filled $d$- and $f$-electron materials closer to a Mott transition for a given Coulomb repulsion, yet drive non-half-filled materials away from a Mott transition while also increasing the correlation within their metallic phases. The consequences of these effects are perhaps best illustrated in 4$d$ transition metal oxides that have more than a single electron or hole in their 4$d$ shells. \cite{Koster_Beasley_RMP_2012,Frandkin_Mackenzie_ARCMP_2010,Dang_Millis_PRB_2015, Han_Millis_PRB_2016} Unlike their rhodate counterparts, which possess a single hole in their shells, many ruthenates and molybdenates exhibit substantial mass enhancements,\cite{Ikeda_JPSJ_2000} unexpected Mott Insulator transitions,\cite{Liebsch_PRL_2007,Gorelov_PRL_2010,Sutter_Chang_NC_2017} novel quantum phase transitions,\cite{Grigera_Science_2001} and even superconducting phases\cite{Mackenzie_RMP_2003} -- all of which may be attributed to Hund's physics.

Despite both the prevalence and importance of Hund's effects, they remain a challenge to describe. Most analytical and numerical treatments revolve around solving a multi-band Hubbard model, most often the Hubbard-Kanamori (HK) model,\cite{Kanamori_PTP_1963} containing a mixture of kinetic, Coulomb $U$, and Hund's $J$ terms. Although analytical studies have been performed,\cite{Roth_PR_1966,LyonCaen_Cyrot_JPC_1975,Khomskii_SolidStateComm_1973,Kugel_Khomskii_Sov_1982,Ishihara_PRB_1997} just as in the case of the single-band Hubbard model containing a repulsive $U$ term,\cite{LeBlanc_Gull_PRX_2015,Zheng_Science_2017} accurate treatments of these models necessitate methods capable of treating strong correlation non-perturbatively. However, because these models possess significantly larger state spaces and involve additional pair-hopping and Hund's exchange terms, they are often even more difficult to treat than the Hubbard model. 

Due to the complicated interactions involved, there is no general analytical solution for these problems. Thus, numerical treatments are in high demand. To date, most numerical studies of multi-band models have employed Dynamical Mean Field Theory (DMFT)\cite{Georges_RevModPhys_1996,Metzner_Vollhardt_PRL_1989} either on its own or in combination with Density Functional Theory (DFT)\cite{Anisimov_JPhysCondMat_1997} because of DMFT's ability to treat band and atomic effects on equal footing by self-consistently solving an impurity problem within a larger bath. DMFT has been very successful at mapping out multi-band phase diagrams at finite temperatures.\cite{Gorelov_PRL_2010,Han_Millis_PRL_2018,Han_Millis_PRB_2016,Dang_Millis_PRB_2015,Dang_PRL_2015,Werner_Andrew_PRL_2008,Inaba_PRB_2005} Nevertheless, DMFT is fundamentally limited by the accuracy and scaling of its impurity model solver. Some DMFT studies rely upon exact diagonalization (ED) to solve their impurity models, yet the computational cost of ED grows exponentially with the number of bands involved, thus thwarting its application to many-band models. Some DMFT algorithms employ continuous-time quantum Monte Carlo (CTQMC)\cite{Rubtsov_PRB_2005} to solve their impurity models. CTQMC can solve larger impurity models than ED, but is still hampered by the sign problem, an exponential decrease in the signal to noise ratio observed in stochastic simulations,\cite{Loh_PRB_1990} in certain parameter regimes and low temperature calculations remain difficult.\cite{Gull_RMP_2011} A method that can accurately  simulate larger system sizes at lower temperatures is thus in need.

One suite of techniques particularly well-suited for studying the large state spaces inherent to multi-band models are quantum Monte Carlo (QMC) techniques.\cite{Foulkes_RMP_2001,Motta_Wiley_2018} Both finite temperature QMC methods, including CTQMC\cite{Gull_RMP_2011} and Hirsch-Fye QMC\cite{Hirsch_PRL_1986} algorithms that have been employed as impurity solvers within DMFT, and ground state\cite{Motome_Imada_JPSJ_1997,Motome_JPSJ_1998} QMC algorithms have been developed and applied to the multi-band Hubbard model. Nonetheless, the Hund's terms of the HK Hamiltonian have posed challenges for all of these methods. This is because Hund's terms are not readily expressed as products of density operators and are therefore not readily amenable to standard QMC transformations. Straightforward decoupling of the exchange and pair hopping terms leads to a severe sign problem.\cite{Held_Vollhardt_EPJB_1998} Attempts have therefore been made to simplify the Hund's contribution to the Hamiltonian to make it more palatable to QMC methods by constraining its direction to the z-axis,\cite{Held_Vollhardt_EPJB_1998,Han_PRB_1998} but such treatments sometimes fail to properly capture the model's expected physics. Several Hund's-specific transformations have been proposed, including a discrete transformation by Aoki \cite{Sakai_Aoki_PRB_2004,Sakai_Aoki_PRB_2006} and a continuous transformation by Imada. \cite{Motome_Imada_JPSJ_1997,Motome_JPSJ_1998} Nevertheless, these transformations ultimately do not eliminate the sign problem and are limited to parameter regimes with only high signal to noise ratios. These parameter constraints obscure our fundamental understanding of multi-band physics. 

In this paper, we present an Auxiliary Field Quantum Monte Carlo (AFQMC) framework especially suited for the study of ground state multi-band Hubbard models and demonstrate its accuracy over a range of realistic parameters using different signal-preserving approximations and trial wave functions. Key to our approach is the strategic use of two forms of both the continuous and discrete Hubbard-Stratonovich (HS) Transformations to decouple the Hund's term: a charge decomposition for negative values of the Hund's coupling parameter, and a spin decomposition for positive values of the Hund's coupling parameter. We also employ an unconventional form of importance sampling in which we shift propagators instead of auxiliary fields so as to enable importance sampling of discrete transformed propagators. Unlike previous works, we furthermore utilize flexible Generalized Hartree-Fock (GHF) trial wave functions combined with the constrained path and phaseless approximations to tame the sign and phase problems, respectively. Altogether, we find that these improvements yield promising results for a variety of HK model benchmarks. Although the algorithm presented is designed for the ground state, it can easily be adapted for use in finite temperature methods.\cite{Liu_JCTC_2018,Zhang_PRL_1999} Our algorithm therefore paves the way to the high accuracy modeling of the low temperature physics of a wide range of multi-band models and materials over a dramatically larger portion of the phase diagram.       

The remainder of the paper is organized as follows. In Section \ref{method}, we outline the HK model, summarize the key features of the AFQMC method, and describe how the conventional AFQMC technique may be modified to best accommodate the HK Hamiltonian. In Section \ref{results}, we then present benchmarks of our method's performance within different parameter regimes, using different trial wave functions, and employing different approximations on two- and three-band HK models for which ED results may be obtained. Towards the end of this section, we also demonstrate the accuracy with which our techniques can predict the charge gaps and magnetic ordering of two-dimensional lattice models far beyond the reach of most other techniques. We conclude with a discussion of the broader implications of this work and future directions in Section \ref{conclusions}.

% --------------------------------------------------------------------------------------
\section{Methods}
\label{method}
\subsection{Hubbard Kanamori Model Hamiltonian}
The HK model is a multi-band version of the Hubbard model designed to account for the competition between the spin and band degrees of freedom observed in the physics of $d$- and $f$- electron material.\cite{Kanamori_PTP_1963,Georges_AnnRev_2013} In order to accomplish this, the model includes not only standard Hubbard on-site density-density interactions, but also inter-band density, exchange, and pair hopping terms. The full HK Hamiltonian, written as general as possible, reads
\begin{equation}
\hat{H} \equiv \hat{H}_{1} + \hat{H}_{2} \equiv \hat{H}_{1} + \hat{H}_{U} + \hat{H}_{J}, 
\label{Hamiltonian}
\end{equation}
where 
\begin{equation}
\hat{H}_{1} = \sum_{i m \sigma}\sum_{j m^\prime \sigma^\prime} t_{im,jm^\prime}^{\sigma\sigma^\prime}\hat{c}_{im\sigma}^{\dagger}\hat{c}_{jm^\prime \sigma^\prime}, 
\label{HOne_Term} 
\end{equation}
\begin{eqnarray}
\hat{H}_{U} &=& 
 \sum_{i,m} U_{im} \hat{n}_{im\uparrow}\hat{n}_{im\downarrow} \nonumber + \sum_{i, m \neq m^\prime} U^{\prime}_{imm'} \hat{n}_{im \uparrow} \hat{n}_{im^\prime \downarrow } \nonumber \\
&+& \sum_{i,m<m^\prime, \sigma } (U^{\prime}_{imm'} - J_{imm'}) \hat{n}_{im \sigma}\hat{n}_{im^\prime \sigma},
\label{HU_Term} 
\end{eqnarray}
and
\begin{equation}
\begin{split}
\hat{H}_{J}
= \sum_{i, m \neq m^\prime}  J_{imm^\prime} &(\hat{c}_{im\uparrow}^{\dagger} \hat{c}_{im^\prime\downarrow}^{\dagger} \hat{c}_{im\downarrow} \hat{c}_{im^\prime\uparrow}  \\
&+\hat{c}_{im\uparrow}^{\dagger}\hat{c}_{im\downarrow}^{\dagger}\hat{c}_{im^\prime\downarrow}\hat{c}_{im^\prime\uparrow} + H.c.).
\end{split}
\label{HJ_Term}
\end{equation}
In the above,  $\hat{c}_{im\sigma}^{\dagger}$($\hat{c}_{im\sigma}$) creates (annihilates) an electron with spin $\sigma$ in band $m$ at site $i$. $\hat{n}$ denotes the number operator and $\hat{n}_{im\uparrow}$, for example, represents the number of spin-up electrons at site $i$ in band $m$. $\hat{H}_{1}$ contains all one-body contributions to the Hamiltonian, including terms parameterized by the constants $t_{im,jm^\prime}^{\sigma \sigma'}$ that describe spin-orbit coupling and the hopping of electrons in different bands between sites $i$ and $j$. $\hat{H}_{2}$ denotes the collection of all two-body operators. $\hat{H}_{U}$ contains all density-density interactions, including the intraband ($U$) and interband ($U^\prime$) Coulomb interactions, and the $z$- (or Ising) component of the Hund's coupling. In contrast, $\hat{H}_{J}$ contains all of the terms that cannot be written as density-density interactions, which consist of the $x$- and $y$- components (spin-exchange) of the Hund's coupling, ($\hat{c}_{im\uparrow}^{\dagger} \hat{c}_{im^\prime\downarrow}^{\dagger} \hat{c}_{im\downarrow} \hat{c}_{im^\prime\uparrow} + H.c. $),
   as well as the pair-hopping interaction ($\hat{c}_{im\uparrow}^{\dagger}\hat{c}_{im\downarrow}^{\dagger}\hat{c}_{im^\prime\downarrow}\hat{c}_{im^\prime\uparrow} +H.c.$), in which two electrons in a given band transfer as a pair to other bands. $J$ denotes the Hund's coupling constant. Note that our formalism is general and allows for band- and site-dependent $U$, $U$', and $J$ constants. 
   
   %The Hamiltonian is rotationally invariant not only in spin space, but also in real (band) space if the condition $U=U^\prime+2J$ is satisfied, \BR{no idea what is meant by this...which comes from the symmetry preserved between the Coulombic matrix elements for bands in a central field.}

\subsection{Modified Hubbard Kanamori Model Hamiltonian}

In order to facilitate programming and the generalization of this HK Hamiltonian into a form in which all coupling constants are independent, we map the Hamiltonian given by Equations \eqref{Hamiltonian}-\eqref{HJ_Term} into a one-band model whose terms only depend upon their band indices. If we now let $i$ and $j$ denote superindices that combine both lattice site and band information, then  
\begin{equation}
\begin{split}
\hat{H}
&= \hat{H}_{1} + \hat{H}_{2} \\
&= \sum_{ij,\sigma\sigma^\prime} t_{ij}^{\sigma\sigma^\prime} \hat{c}_{i\sigma}^{\dagger}\hat{c}_{j\sigma^\prime}\\
&+ \sum_{i} U^{i} \hat{n}_{i\uparrow} \hat{n}_{i\downarrow}\\
&+ \sum_{i<j} U_{1}^{ij}  (\hat{n}_{i\uparrow} \hat{n}_{j\downarrow}+\hat{n}_{i\downarrow}\hat{n}_{j\uparrow})\\
&+ \sum_{i<j} U_{2}^{ij} (\hat{n}_{i\uparrow}\hat{n}_{j\uparrow}+\hat{n}_{i\downarrow}\hat{n}_{j\downarrow})\\
&+ \sum_{i<j} J^{ij} (\hat{c}_{i\uparrow}^{\dagger}\hat{c}_{j\downarrow}^{\dagger}\hat{c}_{i\downarrow}\hat{c}_{j\uparrow}
                     +\hat{c}_{i\uparrow}^{\dagger}\hat{c}_{i\downarrow}^{\dagger}\hat{c}_{j\downarrow}\hat{c}_{j\uparrow} \\
                     &+\hat{c}_{j\uparrow}^{\dagger}\hat{c}_{i\downarrow}^{\dagger}\hat{c}_{j\downarrow}\hat{c}_{i\uparrow}
                     +\hat{c}_{j\uparrow}^{\dagger}\hat{c}_{j\downarrow}^{\dagger}\hat{c}_{i\downarrow}\hat{c}_{i\uparrow}).
\end{split}
\label{Reformed_Hamiltonian}
\end{equation}
$t_{ij}^{\sigma\sigma^\prime}$ describes the hopping and spin-orbit coupling between different sites and bands. In keeping with the $\sum_{i, m<m'}$ and $\sum_{i, m\neq m'}$ summations in Equations \eqref{HU_Term} and \eqref{HJ_Term}, $\sum_{i<j}$ only sums over index combinations that reference different bands on the same site. In this modified HK Model, the $U$ term describes density-density interactions only between electrons with opposite spins in the same band, the $U_{1}$ term describes interactions between electrons with opposite spins in different bands on the same site, the $U_{2}$ term describes interactions between electrons with parallel spins in different bands on the same site, and the $J$ term describes spin-exchange and pair hopping interactions on the same site. Thus, in going from Equations \eqref{HU_Term} and \eqref{HJ_Term} to Equation \eqref{Reformed_Hamiltonian}, the original $U'$ term has become the $U_{1}$ term, the original $(U'-J)$ term has become the $U_{2}$ term, and the $J$ term has been re-expressed. Using Equation \eqref{Reformed_Hamiltonian}, we map a multi-band model into a single-band model in which the number of lattice sites has been enlarged into the number of bands. Since there is no explicit index in our model, we can deal with any number of bands as long as the mapping is done correctly.

\subsection{Overview of AFQMC}

In the remainder of this work, AFQMC will be employed to obtain accurate numerical solutions to the HK Model. AFQMC is a quantum many-body method that solves the ground state Schrodinger Equation by randomly sampling an overcomplete space of non-orthogonal Slater determinants\cite{Zhang_Gubernatis_PRB_1997,Zhang_Book_2003,Zhang_Book_2013} and has consistently been demonstrated to be among the most accurate of modern many-body methods for modeling the Hubbard model over a wide range of parameter regimes. \cite{LeBlanc_Gull_PRX_2015,Zheng_Science_2017,Chang_PRL_2010,Chang_PRB_2008,Shi_PRB_2013} At its heart, AFQMC is an imaginary-time projection quantum Monte Carlo technique that applies a projection operator, $e^{-\beta \hat{H}}$, onto an initial wave function, $|\Psi_{I} \rangle$,
\begin{equation}
|\Psi_{0} \rangle \propto \lim_{\beta \rightarrow \infty} \left(e^{-\beta \hat{H}}\right) | \Psi_{I} \rangle.  
\end{equation}
In the limit of infinite imaginary projection time $(\beta \rightarrow \infty)$, it converges to the ground state wave function, $|\Psi_{0}\rangle$, as long as the initial wave function is not orthogonal to the ground state wave function. Because the projection operator cannot be evaluated for large values of $\beta$, it is discretized into $n = \beta/\Delta\tau$ smaller time slices for which it can be evaluated 
\begin{equation}
|\Psi_{0} \rangle \propto \lim_{n \rightarrow \infty} \left(e^{-\Delta \tau \hat{H}}\right) ^{n} | \Psi_{I} \rangle, 
\end{equation}
and the projection is carried out iteratively as follows
\begin{equation}
|\Psi^{(n+1)} \rangle = e^{-\Delta \tau \hat{H}} | \Psi^{(n)} \rangle. 
\label{iteration}
\end{equation}
For sufficiently small $\Delta \tau$, the projection operator may be factored into one- and two-body pieces via Suzuki-Trotter Factorization\cite{Suzuki_ProgTheorPhys_1976,Trotter_PAMS_1959} 
\begin{equation}
e^{-\Delta\tau \hat{H}} \approx e^{-\Delta\tau \hat{H}_{1}/2} e^{-\Delta\tau \hat{H}_{2}} e^{-\Delta\tau \hat{H}_{1}/2}.
\label{Projection_Equation}
\end{equation}
The two-body propagator may be further decomposed into the four terms given in Equation \eqref{Reformed_Hamiltonian}
\begin{equation}
\begin{split}
e^{-\Delta\tau \hat{H}_{2}} 
&\approx e^{-\Delta\tau \hat{H}_{U}}
e^{-\Delta\tau \hat{H}_{U_1}}e^{-\Delta\tau \hat{H}_{U_2}}e^{-\Delta\tau \hat{H}_{J}}\\
&=e^{-\Delta\tau \sum\limits_{i} U_i \hat{n}_{i\uparrow} \hat{n}_{i\downarrow}}
e^{-\Delta\tau \sum\limits_{i<j} U_1^{ij} (\hat{n}_{i\uparrow} \hat{n}_{j\downarrow}+\hat{n}_{i\downarrow} \hat{n}_{j\uparrow})}\\
&e^{-\Delta\tau \sum\limits_{i<j} U_2^{ij} (\hat{n}_{i\uparrow} \hat{n}_{j\uparrow}+\hat{n}_{i\downarrow} \hat{n}_{j\downarrow})}\\
&e^{-\Delta\tau \sum\limits_{i<j}J^{ij}(\hat{c}_{i\uparrow}^{\dagger}\hat{c}_{j\downarrow}^{\dagger}\hat{c}_{i\downarrow}\hat{c}_{j\uparrow}
                     +\hat{c}_{i\uparrow}^{\dagger}\hat{c}_{i\downarrow}^{\dagger}\hat{c}_{j\downarrow}\hat{c}_{j\uparrow} 
                     +H.c.)}.
\end{split}
\label{two-body propagator}
\end{equation}
A time step extrapolation is needed to make sure the Trotter error is negligible in the Monte Carlo simulation. 

\subsection{Hubbard-Stratonovich Transformation of the Modified Hubbard Kanamori Hamiltonian \label{HSTransform}}
According to Thouless's Theorem, \cite{Thouless_NP_1960} acting the exponential of a one-body operator on a determinant results in another determinant, reducing the process of projecting a one-body operator onto the wave function into standard matrix multiplication. Nevertheless, no such theorem applies to exponentials of two-body operators, which necessitates re-expressing these operators into integrals over one-body operators using the so-called Hubbard-Stratonovich transformation. \cite{Hubbard_PRL_1959}

In order to transform the two-body propagator given by Equation \eqref{two-body propagator}, both discrete\cite{Hirsch_PRB_1983,Gubernatis_Werner} and continuous\cite{Buendia_PRB_1986} HS transformations need to be performed. The $U$, $U_{1}$, and $U_{2}$ terms are products of density operators, much like the conventional Hubbard $U$ term, and may therefore be decomposed using discrete transformations. For $\alpha < 0$, where $\alpha$ may be denote $U$, $U^{1}$, or $U^{2}$, it is usually better to use the discrete charge decomposition
\begin{equation}
e^{-\Delta\tau\alpha \hat{n}_{1}\hat{n}_{2}}=e^{-\Delta\tau\alpha (\hat{n}_{1}+\hat{n}_{2}-1)/2}\sum_{x=\pm{1}}\frac{1}{2}e^{\gamma x (\hat{n}_{1}+ \hat{n}_{2}-1)},
\label{density_charge}
\end{equation}
where
$\cosh(\gamma)=e^{-\Delta\tau\alpha/2}$, while for $\alpha > 0 $, it is usually better to use the spin decomposition
\begin{equation}
e^{-\Delta\tau\alpha \hat{n}_{1}\hat{n}_{2}}=e^{-\Delta\tau\alpha (\hat{n}_{1}+\hat{n}_{2})/2}\sum_{x=\pm{1}}\frac{1}{2}e^{\gamma x (\hat{n}_{1}-\hat{n}_{2})},
\label{density_spin}
\end{equation}
where $\cosh(\gamma)=e^{\Delta\tau\alpha/2}$. In both Equations \eqref{density_charge} and \eqref{density_spin}, $x$ represents the namesake auxiliary field that may assume the discrete values of $+1$ or $-1$. For the subsequent discussion, note that the charge decomposition is so named because it produces a one-body propagator involving the sum of $\hat{n}_{1}+\hat{n}_{2}$, which would be equivalent to the charge on a site if $1$ represented an up and $2$ a down spin on that site. Along similar lines, the spin decomposition is so named because it involves the difference between $\hat{n}_{1}$ and $\hat{n}_{2}$, which would represent the spin on a site under the same assumptions. 

Because $\hat{H}_{J}$ contains terms that are not simple products of density operators, decomposing it is a much more challenging task. Past attempts have either neglected or simplified $\hat{H}_{J}$.\cite{Held_Vollhardt_EPJB_1998,Han_PRB_1998} Several techniques have employed exact decompositions,\cite{Motome_Imada_JPSJ_1997,Sakai_Aoki_PRB_2006,Sakai_Aoki_PRB_2004} but all such decompositions are accompanied by a sign problem that thwarts explorations of wide swaths of the phase diagram. Unlike these past attempts, in the following, we define a unique decomposition that can be employed in both continuous and discrete transformations, and accompany it by importance sampling that first mitigates and the constrained path and phaseless approximations that eliminate the sign and phase problems. As part of our decomposition of $e^{-\Delta \tau \hat{H}_{J}}$, we first re-expressed $\hat{H}_{J}$ in terms of squares of one-body operators. Let \begin{equation}
\hat{\rho}_{ij} \equiv \sum_{\sigma}(\hat{c}_{i\sigma}^{\dagger}\hat{c}_{j\sigma} + \hat{c}_{j\sigma}^{\dagger}\hat{c}_{i\sigma}). 
\end{equation} 
Then,  
\begin{equation}
\hat{\rho}_{ij}^{2} = \sum_{\sigma\sigma^\prime}(\hat{c}_{i\sigma}^{\dagger}\hat{c}_{j\sigma} + \hat{c}_{j\sigma}^{\dagger}\hat{c}_{i\sigma}) (\hat{c}_{i\sigma^\prime}^{\dagger}\hat{c}_{j\sigma^\prime} + \hat{c}_{j\sigma^\prime}^{\dagger}\hat{c}_{i\sigma^\prime}), 
\end{equation}
and $\hat{H}_{J}$ may be re-expressed as (see the Supplemental Materials for more details)
\begin{equation}
\begin{split}
\hat{H}_{J}
&=\sum_{i<j} J^{ij}(\hat{c}_{i\uparrow}^{\dagger}\hat{c}_{j\downarrow}^{\dagger}\hat{c}_{i\downarrow}\hat{c}_{j\uparrow}+\hat{c}_{i\uparrow}^{\dagger}\hat{c}_{i\downarrow}^{\dagger}\hat{c}_{j\downarrow}\hat{c}_{j\uparrow}+ H.c.) \\
&=\sum_{i<j} \frac{J^{ij}}{2} [\hat{\rho}_{ij}^{2} - \sum_{\sigma} (\hat{n}_{i\sigma}+\hat{n}_{j\sigma} - \hat{n}_{i\sigma}\hat{n}_{j\sigma}-\hat{n}_{j\sigma}\hat{n}_{i\sigma})] \\
&=\sum_{i<j} \frac{J^{ij}}{2}\hat{\rho}_{ij}^{2} - \sum_{i<j, \sigma} \frac{J^{ij}}{2} (\hat{n}_{i\sigma}+\hat{n}_{j\sigma}) +\sum_{i<j, \sigma} J^{ij} \hat{n}_{i\sigma}\hat{n}_{j\sigma}.
\end{split}
\label{transformation_1}
\end{equation}
The second term of Equation \eqref{transformation_1} consists of one-body operators and can be combined with the other one-body operators into $\hat{H}_{1}$. The third term consists of a product of density operators and can therefore be transformed according to either Equations \eqref{density_charge} or \eqref{density_spin}. The first term, however, consists of a square that cannot be resolved into products of density operators. In order to decouple this two-body term, a continuous HS transformation must be employed. In general, the continuous HS transformation may be written as
\begin{equation}
e^{-\Delta \tau \hat{A}^{2}/2} = \int dx \frac{1}{\sqrt{2\pi}} e^{-x^{2}/2} e^{x\sqrt{-\Delta \tau}\hat{A}}
\label{continuous_HS_transformation}, 
\end{equation}
where $\hat{A}$ represents any one-body operator and $x$ denotes an auxiliary field, as before. Letting $\hat{A} \equiv \hat{\rho}_{ij}$, it follows that the most obvious way to transform the exponential formed from the first term of Equation \eqref{transformation_1} is using the charge decomposition
\begin{equation}
\begin{split}
& e^{-\Delta\tau  \sum\limits_{i<j} 
\frac{J^{ij} }{2}[\sum\limits_{\sigma}(\hat{c}_{i\sigma}^{\dagger}\hat{c}_{j\sigma}+\hat{c}_{j\sigma}^{\dagger}\hat{c}_{i\sigma})]^{2}} \\
&=
\prod_{i<j}\int dx_{ij} \frac{1}{\sqrt{2\pi}} e^{- x_{ij}^{2}/2} e^{x_{ij}\sqrt{-\Delta\tau J^{ij}} [\sum\limits_{\sigma}(\hat{c}_{i\sigma}^{\dagger}\hat{c}_{j\sigma}+\hat{c}_{j\sigma}^{\dagger}\hat{c}_{i\sigma})]}.
\end{split}
\label{transformation_charge}
\end{equation}
As long as $J^{ij}<0$ for all $i,j$, all of the propagators produced by this transformation will be real, as is desirable within AFQMC simulations. However, if any of the $J^{ij}$ are greater than 0, $\sqrt{-\Delta\tau J^{ij}}$ will be complex resulting in a complex propagator that immediately introduces a complex phase into simulations. To prevent complexity from being introduced into the operators, in certain cases, we take a cue from the discrete case and define a continuous spin decomposition that involves the difference between spin up and down operators. Let 
\begin{equation}
\hat{\rho}_{ij} = \sum_{\sigma} \delta _{\sigma}(\hat{c}_{i\sigma}^{\dagger}\hat{c}_{j\sigma} + \hat{c}_{j\sigma}^{\dagger}\hat{c}_{i\sigma}),
\label{Rho_Equation}
\end{equation}
where $\delta _{\uparrow}=1$ and $\delta _{\downarrow}=-1$, then (see the Supplemental Materials for further details)
\begin{equation}
\hat{\rho}_{ij}^{2} = \sum_{\sigma\sigma^\prime}
\delta _{\sigma}\delta _{\sigma^\prime}(\hat{c}_{i\sigma}^{\dagger}\hat{c}_{j\sigma} + \hat{c}_{j\sigma}^{\dagger}\hat{c}_{i\sigma}) (\hat{c}_{i\sigma^\prime}^{\dagger}\hat{c}_{j\sigma^\prime} + \hat{c}_{j\sigma^\prime}^{\dagger}\hat{c}_{i\sigma^\prime}).
\label{Rho_Equation_Squared}
\end{equation}
Using this to re-express $\hat{H}_{J}$, we have 
\begin{equation}
\begin{split}
\hat{H}_{J}
&=\sum_{i<j} J^{ij}(\hat{c}_{i\uparrow}^{\dagger}\hat{c}_{j\downarrow}^{\dagger}\hat{c}_{i\downarrow}\hat{c}_{j\uparrow}+\hat{c}_{i\uparrow}^{\dagger}\hat{c}_{i\downarrow}^{\dagger}\hat{c}_{j\downarrow}\hat{c}_{j\uparrow}+ H .c.) \\
&=\sum_{i<j} - \frac{J^{ij}}{2} [\hat{\rho}_{ij}^{2} - \sum_{\sigma} (\hat{n}_{i\sigma}+\hat{n}_{j\sigma} - \hat{n}_{i\sigma}\hat{n}_{j\sigma}-\hat{n}_{j\sigma}\hat{n}_{i\sigma})] \\
&=\sum_{i<j} -\frac{J^{ij}}{2}\hat{\rho}_{ij}^{2} +\sum_{i<j, \sigma} \frac{J^{ij}}{2} (\hat{n}_{i\sigma}+\hat{n}_{j\sigma}) -\sum_{i<j, \sigma} J^{ij} \hat{n}_{i\sigma}\hat{n}_{j\sigma}.
\end{split}
\label{transformation_3}
\end{equation}
Employing this form for the decomposition, the exponential that stems from the first term of Equation \eqref{transformation_3} may now be transformed to yield
\begin{equation}
\begin{split}
& e^{\Delta\tau  \sum\limits_{i<j} 
\frac{J^{ij} }{2}[\sum\limits_{\sigma}\delta _{\sigma}(\hat{c}_{i\sigma}^{\dagger}\hat{c}_{j\sigma}+\hat{c}_{j\sigma}^{\dagger}\hat{c}_{i\sigma})]^{2}} \\
&=
\prod_{i<j}\int dx_{ij} \frac{1}{\sqrt{2\pi}} e^{- x_{ij}^{2}/2} e^{x_{ij}\sqrt{\Delta\tau J^{ij}} [\sum\limits_{\sigma}\delta _{\sigma}(\hat{c}_{i\sigma}^{\dagger}\hat{c}_{j\sigma}+\hat{c}_{j\sigma}^{\dagger}\hat{c}_{i\sigma})]}, 
\end{split}
\label{transformation_spin}
\end{equation}
which is real for $J^{ij} > 0$. Using the charge decomposition (Equation \eqref{transformation_charge}) when $J^{ij}<0$ and the spin decomposition (Equation \eqref{transformation_spin}) when $J^{ij}>0$ thus completely eliminates complex propagators, easing simulation. In Section \ref{two-band}, we compare the merits of using this mixed decomposition approach to exclusively relying upon the complex charge decomposition on the accuracy of our overall results.    

%The phase problem is a generalization of the sign problem observed in determinant Quantum Monte Carlo simulations of the Hubbard model\cite{Loh_PRB_1990,White_PRB_1989} in which walker weights populate the entire complex plane, resulting in a cancellation of weights that produces a signal to noise ratio that decreases exponentially with increasing $\beta$. \cite{Zhang_Krakauer_PRL_2003,Motta_Wiley_2018}

Inserting the HS transformations defined by Equations \eqref{density_charge}, \eqref{density_spin}, \eqref{transformation_charge}, and \eqref{transformation_spin} into Equations \eqref{Projection_Equation} and \eqref{two-body propagator} and combining terms, one arrives at the final AFQMC expression for the projection operator
\begin{equation}
e^{-\Delta \tau \hat{H}} = \int d{\bf x} p({\bf x}) \hat{B}({\bf x}),
\label{effective_propagation}
\end{equation}
where ${\bf x} = \{x_1, x_2, . . . , x_{N_{F}}
\}$ denotes the set of $N_{F}$ total normally distributed auxiliary
fields sampled at a given time slice, $\hat{B}(x)$ represents the amalgamation of all one-body operators, and $p(x)$ is a combination of all scalar functions of the fields. Example expressions for $\hat{B}(x)$ and $p(x)$ are given in the Supplemental Information. As is clear from Equation \eqref{effective_propagation}, the series of HS Transformations described ultimately maps the original two-body propagator into a weighted integral over one-body propagators that are functions of external auxiliary fields.

\subsection{Sampling in AFQMC}

\subsubsection{The Sampling Process}

One of the most computationally efficient ways of evaluating many dimensional integrals such as that given by Equation \eqref{effective_propagation} is to use Monte Carlo sampling techniques. As described in more detail in previous publications,\cite{Hirsch_PRB_1985, Zhang_Gubernatis_PRB_1997,Zhang_Book_2003,Zhang_Book_2013} if $|\Psi_{I}\rangle$ is represented by a single Slater determinant, after each application of the projection operator, a new Slater determinant will be produced. Thus, if $k$ instances (so-called ``walkers'') are initialized to $|\Psi_{I}\rangle$ and the projection operation given by Equation \eqref{effective_propagation} is applied to each of them by independently sampling sets of fields, then a random walk through the space of non-orthogonal determinants is realized in which the overall wave function at time slice $n$,  $|\Psi^{(n)}\rangle$, is represented by an ensemble of $k$ wave functions $|\psi_{k}^{(n)}\rangle$ with weights $w_{k}^{(n)}$ 
\begin{equation}
| \Psi^{(n)}\rangle = \sum_{k} w_k^{(n)} | \psi_k^{(n)} \rangle .
\label{wave function}
\end{equation}
Here, the $w_{k}^{(n)}$ consist of the products of numbers accumulated over all time slices by walker $k$, which can be a complex number. 

Ground state observables at each time slice, such as the energies reported below, may then be computed by evaluating the mixed estimator\cite{Foulkes_RMP_2001} over the ensemble
\begin{eqnarray}
\langle \hat{A} \rangle_{mix} &=& \frac{\langle \Psi_{T} | \hat{A} | \Psi^{(n)} \rangle}{\langle \Psi_{T} | \Psi^{(n)} \rangle} \nonumber \\
&=& \frac{ \sum_{k} w_{k}^{(n)} \langle \Psi_{T} | \hat{A} | \psi_{k}^{(n)} \rangle}{ \sum_{k} w_{k}^{(n)} \langle \Psi_{T} | \psi_{k}^{(n)} \rangle} ,
\label{Mixed}
\end{eqnarray}
where $|\Psi_{T} \rangle$ denotes a trial wave function that approximates the true ground state wave function. To facilitate the evaluation of the mixed estimator, it is common to introduce the local energy
\begin{equation}
E_{L}[\Psi_{T}, \Phi] \equiv \frac{\langle \Psi_{T}| \hat{H} | \Phi \rangle}{\langle \Psi_{T} | \Phi \rangle}, 
\end{equation}
such that Equation \eqref{Mixed} may be simplified to
\begin{equation}
\langle \hat{A} \rangle_{mix} = \frac{ \sum_{k} w_{k}^{(n)}  \langle \Psi_{T} | \psi_{k}^{(n)} \rangle E_{L}[\Psi_{T}, \psi_{k}^{(n)}]}{ \sum_{k} w_{k}^{(n)} \langle \Psi_{T} | \psi_{k}^{(n)} \rangle}. 
\label{Mixed_Again}
\end{equation}
After a sufficiently large number of time slices such that $|\Psi^{(n)}\rangle$ approaches the ground state, final estimates of $\langle \hat{A}\rangle$ may be obtained by averaging over each time slice expectation value.

A population control procedure \cite{Calandra_Sorella_PRB_1998} is needed during the random walk. During this procedure, walkers with larger weights are replicated and those with smaller weights are eliminated probabilistically. The weight used in population control is 
\begin{equation}
W^{(n)}_{k} = w_{k}^{(n)} \langle \Psi_{T} | \psi_{k}^{(n)} \rangle .
\label{weight}
\end{equation}
When there is a sign or phase problem, $ W^{(n)}_{k} $ may become negative or complex. As described in Section (\ref{cpa}) and Section (\ref{pha}), $ W^{(n)}_{k} $ is always positive or zero if the constrained path or phaseless approximations are employed.  

\subsubsection{The Sign and Phase Problems}
Unfortunately, the ``free'' projection process just described is typically beset by either the sign\cite{Loh_PRB_1990,Ceperley1991} or phase problems.\cite{Zhang_Krakauer_PRL_2003} These problems fundamentally stem from the fact that observables computed using a single Slater determinant, $|\Psi\rangle$, remain invariant to arbitrary rotations, $e^{i\theta} |\Psi \rangle$, of that determinant, where $\theta$ is a phase angle. Consequently, during the course of an AFQMC simulation involving complex propagators, walkers may accumulate infinitely many possible phases (as there are infinitely many possible phase angles, $\theta \in [0, 2\pi)$), resulting in infinitely many possible determinants. Since these phases are directly multiplied into the walker weights of Equations \eqref{Mixed} and \eqref{Mixed_Again}, after many iterations, the walker weights end up populating the entire complex plane and many of the terms summed to compute weighted averages of observables cancel one another out. This cancellation leads to an exponential decline in observable signal to noise ratios that manifests as infinite variances\cite{Shi_PRB_2013} called the phase problem. If transformations that preclude propagators from becoming complex are employed as described above, positive and negative versions of each determinant may still be generated, resulting in a somewhat less pernicious cancellation of positive and negative weights termed the sign problem. If left unchecked, the sign and phase problems render obtaining meaningful observable averages nearly impossible, thwarting AFQMC simulations. We therefore mitigate these problems using a combination of background subtraction, importance sampling, and either the constrained path (for the sign problem) or phaseless (for the phase problem) approximations.

\subsubsection{Background Subtraction}

One of the simplest ways of reducing variances within AFQMC is via background subtraction.\cite{Purwanto_PRA_2005} As part of background subtraction, the two-body portion of a Hamiltonian is rewritten so that a mean field average is subtracted from each one-body operator. Thus, if the original two-body operator may be written as a square such that $\hat{V} = -\frac{1}{2} \sum_{i} \hat{v}_{i}^{2}$ to make it amenable to a HS Transformation, as part of background subtraction, it would be re-expressed as 
\begin{equation}
\hat{V} = -\frac{1}{2} \sum_{i} \left(\hat{v}_{i} - \langle \hat{v}_{i} \rangle \right)^{2} - \sum_{i} \hat{v}_{i} \langle \hat{v}_{i} \rangle + \frac{1}{2} \sum_{i} \langle \hat{v}_{i} \rangle^{2}, 
\end{equation}
where $\langle \hat{v}_{i} \rangle$ denotes the mean field average of the operator $\hat{v}_{i}$ (see the Supplemental Materials for more details on how this mean field average is obtained). Because the modified $\hat{v}_{i} - \langle \hat{v}_{i} \rangle$ operator will be smaller in magnitude than the bare $\hat{v}_{i}$ operator, background subtraction reduces the variance involved in AFQMC simulations. In this work, we perform background subtraction on the only term in the Hamiltonian that is not a product of on-site densities, the $\frac{J^{ij}}{2}\hat{\rho}_{ij}^{2}$ term of Equation \eqref{transformation_1} or the $-\frac{J^{ij}}{2}\hat{\rho}_{ij}^{2}$ term of Equation \eqref{transformation_3}, yielding
\begin{eqnarray}
\sum_{i<j} \frac{J^{ij}}{2} \hat{\rho}_{ij}^{2}
&=& \sum_{i<j} \frac{J^{ij}}{2} (\hat{\rho}_{ij}-\langle \hat{\rho}_{ij} \rangle)^{2}
-\sum_{i<j} \frac{J^{ij}}{2} \langle \hat{\rho}_{ij} \rangle^{2} \nonumber \\
&+& \sum_{i<j}J^{ij}\langle \hat{\rho}_{ij} \rangle \hat{\rho}_{ij}
\end{eqnarray} 
and
\begin{eqnarray}
\sum_{i<j} -\frac{J^{ij}}{2} \hat{\rho}_{ij}^{2}
&=&\sum_{i<j} -\frac{J^{ij}}{2} (\hat{\rho}_{ij}-\langle \hat{\rho}_{ij} \rangle)^{2} +\sum_{i<j} \frac{J^{ij}}{2} \langle \hat{\rho}_{ij} \rangle^{2}  \nonumber \\
&-& \sum_{i<j}J^{ij} \langle \hat{\rho}_{ij} \rangle \hat{\rho}_{ij}, 
\end{eqnarray}
respectively. 

\subsubsection{Importance Sampling}
In order to further reduce the variance of walker weights and to make our simulations more amenable to the constrained path and phaseless approximations, we additionally perform importance sampling, which aims to shift the center of the distribution from which we sample our auxiliary fields so that the most important fields are sampled more frequently. The conventional way of performing importance sampling in AFQMC simulations is by introducing a force bias that shifts each sampled field by an amount dependent upon the operator being transformed and the current walker wave function.\cite{Purwanto_PRE_2004,Zhang_Krakauer_PRL_2003,Rom_Neuhauser_CPL_1997,Shi_Zhang_PRA_2015} Because we utilize a mixture of discrete and continuous transformations and force bias importance sampling is only applicable to continuous transformations, in this work, we employ a formally equivalent strategy in which we shift \emph{the propagators instead of the auxiliary fields}. 

For continuous HS Transformations, this may be accomplished by shifting the operator $\hat{A}$ by $\langle \hat{A} \rangle$ in Equation \eqref{continuous_HS_transformation}
\begin{equation}
\begin{split}
e^{-\Delta \tau \hat{A}^{2}/2} 
&= \int dx \frac{1}{\sqrt{2\pi}} e^{-x^{2}/2} e^{x\sqrt{-\Delta \tau}\hat{A}} \\
&= \int dx \frac{1}{\sqrt{2\pi}} e^{-x^{2}/2} e^{x\sqrt{-\Delta \tau}\langle \hat{A} \rangle} e^{x\sqrt{-\Delta \tau}(\hat{A}-\langle \hat{A} \rangle)},  
\label{importance_sampling_continuous}
\end{split}
\end{equation}
where $\langle \hat{A} \rangle$ is the mixed estimator of $\hat{A}$
\begin{equation}
\langle \hat{A} \rangle
\equiv \frac{\langle \Psi_T | \hat{A} | \psi_k^{(n)} \rangle }{\langle \Psi_T | \psi_k^{(n)} \rangle }. 
\end{equation}
If we define the dynamic force as $F \equiv \sqrt{-\Delta \tau}\langle \hat{A} \rangle$, then Equation \eqref{importance_sampling_continuous} may be re-expressed as
\begin{equation}
\begin{split}
e^{-\Delta \tau \hat{A}^{2}/2} 
&= \int dx \frac{1}{\sqrt{2\pi}} e^{-x^{2}/2} e^{xF} e^{x\sqrt{-\Delta \tau}\hat{A}-xF} \\
&= \int dx \frac{1}{\sqrt{2\pi}} e^{-(x-F)^{2}/2} e^{\frac{1}{2}F^2}e^{x\sqrt{-\Delta \tau}\hat{A}-xF} \\
&= \int dx \frac{1}{\sqrt{2\pi}} e^{-(x-F)^{2}/2} e^{\frac{1}{2}F^2-xF}e^{x\sqrt{-\Delta \tau}\hat{A}}
\label{importance_sampling_continuous-2}. 
\end{split}
\end{equation}
In order to realize this transformation, fields are  sampled from the shifted Gaussian probability density function, $\frac{1}{\sqrt{2\pi}} e^{-(x-F)^{2}/2}$, and the propagator $e^{x\sqrt{-\Delta \tau} \hat{A}}$ is applied with weight $e^{\frac{1}{2}F^{2}-xF}$. The field distributions are now centered around the dynamic force, which can be shown to minimize the variance. If the dynamic force $F$ is a complex number, our auxiliary fields will have the same imaginary part to ensure the $x-F$ is real. Then the probability function $\frac{1}{\sqrt{2\pi}} e^{-(x-F)^{2}/2}$ stays in the real axis, which can be sampled by Monte Carlo.

Shifting the propagator within a discrete transformation proceeds in exactly the same fashion. Comparing Equations \eqref{continuous_HS_transformation} and \eqref{density_spin}, the dynamic force needed to shift the propagator in Equation \eqref{density_spin}, for example, would be $F\equiv \gamma(\langle \hat{n}_{1} \rangle - \langle \hat{n}_{2} \rangle)$, resulting in the transformation
\begin{equation}
\begin{split}
e^{-\Delta\tau\alpha \hat{n}_{1}\hat{n}_{2}}
&=e^{-\Delta\tau\alpha (\hat{n}_{1}+\hat{n}_{2})/2}\sum_{x=\pm{1}}\frac{1}{2}e^{\gamma x (\hat{n}_{1}-\hat{n}_{2})} \\
&= e^{-\Delta\tau\alpha (\hat{n}_{1}+\hat{n}_{2})/2} \sum_{x=\pm{1}}\frac{1}{2} \left( \frac{e^{xF}}{W} \right) W e^{\gamma x (\hat{n}_{1}-\hat{n}_{2})-xF}. \\
%&= \sum_{x=\pm{1}} \frac{1}{2} \left(\frac{e^{xF}}{W}\right) W e^{-xF} e^{(\gamma x-\Delta \tau \alpha/2)\hat{n}_{1}} e^{(-\gamma x - \Delta \tau \alpha/2) \hat{n}_{2}} 
\label{importance_sampling-discrete}
\end{split}
\end{equation}
As in the continuous case, in order to realize this transformation, fields are now sampled from a shifted probability density function, $e^{xF}/W$, where $W$ is the normalization factor, $W=e^{xF} + e^{-xF}$, and the propagator $e^{(-\Delta\tau\alpha/2 + \gamma x)\hat{n}_1}e^{(-\Delta\tau\alpha/2 - \gamma x)\hat{n}_2}$ is applied with weight $\frac{1}{2}We^{-xF}$. A shifted transformation may similarly be constructed for the discrete charge decomposition given by Equation \eqref{density_charge}. Propagators that include background subtraction may be shifted by simply replacing $\hat{A}$ with $\hat{A}-\langle \hat{A} \rangle$ in Equations \eqref{importance_sampling_continuous} and \eqref{importance_sampling_continuous-2} above (see the Supplemental Materials). 

It can readily be proven that shifting auxiliary fields is equivalent to shifting propagators.\cite{Rom_Neuhauser_CPL_1997,Rom_JCP_1998,Shi_Zhang_PRA_2015} Shifting propagators therefore entails a convenient way of introducing importance sampling when discrete transformations are involved. Overall, the importance sampled propagation produces the same observable averages as free propagation, but favors the sampling of determinants with larger overlaps with the trial wave function and suppressing the sampling of determinants with no overlap.  

%\BR{I don't know that we need to include what follows below...does this really tell us anything?
%With combination with importance sampling, we have a modified propagation path
%\begin{equation}
%| \Psi^{(n+1)}\rangle = \int dx p^*(x,F) p(x)\frac{1}{p^*(x,F)} \hat{B}(x) | \Psi^{(n)} \rangle
%\end{equation}
%where only the probability density function is modified to $p^*(x,F)$.
%%the probability distribution for $x$ vanishes smoothly as $O_T(\hat{B}(x)\phi^{(n)})$ approaches zero, and the constraint is naturally imposed.
%the weight is modified accounting for the constants ($\frac{1}{p^*(x,F)}$) from the propagator and the normalization factor ($W$) from the probability density function.}

%It can be easily proved that the shift of the auxiliary field or the shift of the propagator is the same in nature. But the later strategy can be easily applied to discrete HS transformation and the description of the walker would be easier since there is no need to introduce importance function ($\langle \Psi_T | \psi \rangle $) for walker sampling and propagation doesn't need to be modified. The sampling center is determined by the dynamic force, which is naturally the optimal choice. Later we will use this strategy for the following constrained path approximation and phaseless approximation.
%%As is shown later, the introduction of importance sampling with force bias can guide the walker away from the the nodal surface where the weight $w(\Psi) \approx 0$ as $\bar{x}$ diverges. This is important for preventing the walkers from crossing the node and entering the other degenerate spaces. 

\subsubsection{Constrained Path Approximation}
\label{cpa}

In order to address the sign problem that may emerge when our propagators, $\hat{B}(\vec{x})$, are real, we employ the constrained path approximation. \cite{Zhang_Gubernatis_PRB_1997} Here, we impose this approximation by requiring that all walkers maintain a positive overlap with the trial wave function after each propagation step 
\begin{equation}
w_k^{(n)}\langle \Psi_T | \psi_k^{(n)} \rangle >0. 
\label{overlap}
\end{equation}
As in typical constrained path implementations, walkers with negative overlaps with the trial wave function will be killed (have their weights set equal to zero), preventing them from being propagated further. This condition will select for only walkers with positive determinants, eliminating the sign problem. It can be shown that if the trial wave function is the exact ground state wave function, this condition will be exact;\cite{Carlson_PRB_1999} however, since the trial wave function is typically unknown, constraining the propagation path in this way results in a small, but consequential approximation.\cite{Chang_PRB_2016,Shi_PRB_2013} 

\subsubsection{Phaseless Approximation}
\label{pha}

In cases in which our propagators are complex, instead of employing the constrained path approximation, we employ the more general phaseless approximation.\cite{Zhang_Krakauer_PRL_2003,Purwanto_PRA_2005} The phaseless approximation controls the phase problem by projecting  complex walker weights onto the positive real axis according to the equation 
\begin{equation}
\label{cosProjection}
W^{(n)}_{k} = |W^{(n)}_{k}| \times \max (0, \cos(\Delta \theta)),
\end{equation}
where $W^{(n)}_{k}$ is defined in Equation \eqref{weight} and $\Delta \theta$, the phase angle, is defined as 
\begin{equation}
\begin{split}
&\Delta \theta =Arg \left[ \frac{\langle \Psi_T | \hat{B}(x) | \psi_k^{(n)} \rangle }{\langle \Psi_T | \psi_k^{(n)} \rangle } \right] \approx O(Im(xF)).
\end{split}
\end{equation}
The use of the cosine function to project also ensures that the density of the walkers will vanish at the origin. Because this cosine projection does not affect walkers with real weights, in practical implementations, we apply Equation \ref{cosProjection} to realize both the constrained path and phaseless Approximations.  

\subsection{Trial and Initial Wave Functions}
Although AFQMC can readily accommodate multi-determinant trial wave functions, we restrict ourselves to employing single determinant trial wave functions that satisfy certain symmetries\cite{Shi_PRB_2013} such as the free electron (FE), restricted Hartree-Fock (RHF), unrestricted Hartree-Fock (UHF), and generalized Hartree-Fock (GHF) wave functions. RHF wave functions preserve spin symmetry. While RHF and UHF wave functions separately conserve the number spin up and down electrons, GHF only fix the total number of electrons. Details about how these wave functions are generated may be found in the Supplemental Materials. 

As illustrated in what follows, because GHF wave functions do not impose any spin symmetries and are therefore the most flexible of these wave function ansatzes, they enable the fastest AFQMC wave function relaxation to the global energy minimum. Nevertheless, when the number of up and down electrons must be fixed, UHF/RHF wave functions were employed instead. Even though our formalism permits our initial wave functions to differ from our trial wave functions, we take our initial and trial wave functions to be the same, except where otherwise noted. 

\section{Results and Discussion}
\label{results}

\subsection{Two-Band Hubbard Kanamori Model Benchmarks}
\label{two-band}

In order to test the accuracy of our theoretical framework, we began by benchmarking our method against ED results for the one-dimensional, two-band HK Model on $5 \times 1$ and $6 \times 1$ lattices with periodic boundary conditions small enough to diagonalize. For these benchmarks, we simplify the Hamiltonian given by Equations \eqref{Hamiltonian} and \eqref{HOne_Term} so that hopping can only occur between adjacent sites within the same bands and may be described by a single site- and spin-invariant constant $t$, such that
\begin{equation}
\hat{H}_{1}^{'} = -t \sum_{\langle ij \rangle,\sigma }\sum_{m=1}^{2}  \hat{c}_{im\sigma}^{\dagger}\hat{c}_{jm\sigma}. 
\end{equation}
We moreover assume that the parameters are site-invariant, such that U$^{i}$=U, U$_1^{ij}$ = U$_1$, U$_2^{ij}$ = U$_2$, and J$^{ij}$ = J. 
 
\textbf{\begin{table}[htbp]
\footnotesize
\caption{The ground state energy of the two-band, 6$\times$1 HK model with $N_{\uparrow}=N_{\downarrow}=6$ over a range of parameters using ED and AFQMC. All energies and parameters are reported in units of $t$.}
\label{table2}
\begin{ruledtabular}
\begin{tabular}{cccccc}
    U & U$_{1}$ & U$_{2}$ & J & ED & AFQMC \\ 
\hline
    2.0 & 1.5 & 1.0 & 0.5 & -3.773268 & -3.774(3)  \\
    2.0 & 1.5 & 1.0 & 1.0 & -4.234037 & -4.230(6)  \\
    2.0 & 1.5 & 3.0 & 0.5 & 0.758540 & 0.755(4)  \\
    3.0 & 5.0 & 1.0 & 0.5 & 2.460374 & 2.466(5)  \\
    6.0 & 1.5 & 1.0 & 0.5 & 1.496509 & 1.503(6)  \\
\end{tabular}
\end{ruledtabular}
\end{table}}

Table \ref{table2} presents our results for a $6\times 1$ HK model over a representative sampling of parameters at half filling. All of the calculations presented were initialized using 560 walkers and employed FE trial and initial wave functions, except for the $U$=3.0,$U_{1}$=5.0,$U_{2}$=1.0,$J$=0.5 case. In this case, it was found that an RHF trial wave function yielded a lower trial energy and manifested a different spin order (antiferromagnetic (AFM) order between two bands) than the FE solution. Thus, an RHF trial wave function was employed instead. This demonstrates that trial wave functions should first be analyzed to determine whether their global minima exhibit the correct order before using them to guide propagation within AFQMC. Unless otherwise noted, all of the results presented in this section were obtained using a charge decomposition for $J$ and the phaseless approximation to tame the related phase problem that emerges. 

As is clear from the table, AFQMC results, are within $0.01t$ or less of exact results, with the smallest discrepancy occurring for the $U=2.0$, $U_{1}=1.5$ case and the largest occurring for the $U=6.0$ case. In all of these cases, exact results are within two standard derivations of the Monte Carlo results, despite the use of the phaseless approximation. 

\begin{center}
\begin{figure}[htbp]
\includegraphics[width=10.5cm]{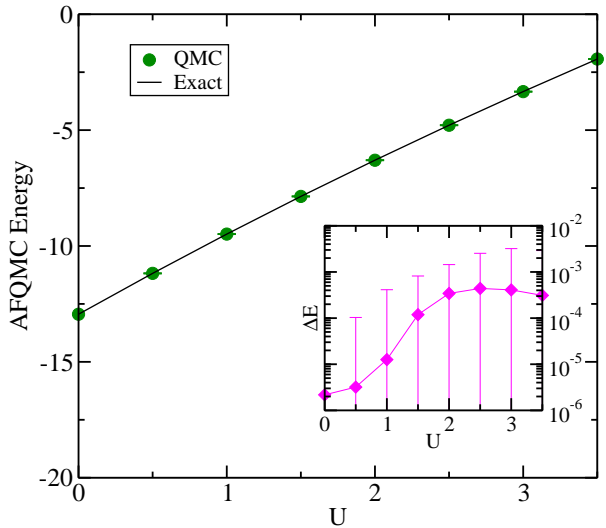}
\captionsetup{font=footnotesize, justification=raggedright, singlelinecheck=false}
\caption{AFQMC ground state energy vs. the density-density parameter $U$ for the two-band, 5$\times$1 HK model using the charge decomposition and FE trial wave functions. Here, all of the other Hamiltonian parameters are held fixed at $t=1$, $U_1$=0, $U_2$=0, and $J=0$ with $N_{\uparrow}=N_{\downarrow}=6$. Relative errors, $\Delta E$, taken with respect to ED result are plotted in the inset for clarity.}
\label{f1}
\end{figure}
\end{center}

\begin{center}
\begin{figure}[htbp]
\includegraphics[width=8.5cm]{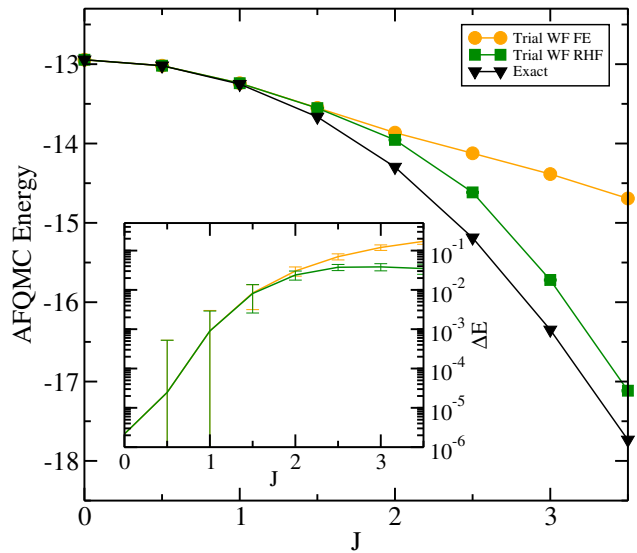}
\captionsetup{font=footnotesize, justification=raggedright, singlelinecheck=false}
\caption{AFQMC ground state energy vs. the Hund's coupling parameter $J$ for the two-band, 5$\times$1 HK model using the charge decomposition and FE/RHF trial wave functions (WF). Here, all of the other Hamiltonian parameters are held fixed at $t=1$, $U=0$, $U_1$=0, and $U_2$=0 with $N_{\uparrow}=N_{\downarrow}=6$. Relative errors, $\Delta E$, taken with respect to ED results are plotted in the inset for clarity.}
\label{f2}
\end{figure}
\end{center}

To pinpoint AFQMC systematic bias, as well as to better understand which regions of the phase diagram are the most challenging for AFQMC, we independently scanned through each of the $U$, $U_{1}$, $U_{2}$, and $J$ parameters holding the others fixed for a $5 \times 1$ HK model. In Figures \ref{f1} and \ref{f2}, we present our scans over $U$ and $J$; figures of our $U_{1}$ and $U_{2}$ scans are presented in the Supplemental Materials.

%%%% begin figure----------------------------------------------
%%%% end figure----------------------------------------------
As shown in Figure \ref{f1}, although the magnitude of the error bars grows with $U$, the relative error remains within 0.1\% to 1\% throughout this range. Similar trends are observed for $U_{1}$ and $U_{2}$. This gives us reason to believe that our method can readily accommodate some of the even larger $U$ values used in studies of strongly correlated materials. Nevertheless, much larger relative errors are observed as $J$ is varied, as depicted in Figure \ref{f2}. This is consistent with previous work, which also implicates the $J$ terms as being most conducive to QMC errors.\cite{Held_Vollhardt_EPJB_1998} Fortunately, for most real materials, $J$ is usually a small fraction of $U$. For small $J$ values, the relative errors are observed to remain less than 1\% and are therefore controllable. 
%%%% begin figure----------------------------------------------
%%%% end figure----------------------------------------------
What may also be gleaned from Figure \ref{f2} is that the quality of the $J>1.5$ energies depends upon the type of trial wave function employed. While free propagation calculations yield results that are independent of the trial wave function, the quality of the constrained path and phaseless approximations fundamentally depend on the accuracy of the trial wave function employed. As depicted in Figure \ref{f2}, the relative errors in the energies produced by FE trial wave functions surpass 10\% and increase with increasing $J$; in contrast, the relative errors produced by RHF trial wave functions not only remain less than 10\%, but plateau as a function of $J$. As $J$ increases, the RHF electron density becomes non-uniform, yielding a lower variational energy than the FE wave function. Figure \ref{f2} thus demonstrates that AFQMC becomes more accurate as trial wave functions better describe the ground state. Note that we also tested UHF and GHF wave functions, which all converged to the same states as RHF wave functions.
%%%% begin figure----------------------------------------------
%\begin{center}
\begin{figure}[htp]
\includegraphics[width=9cm]{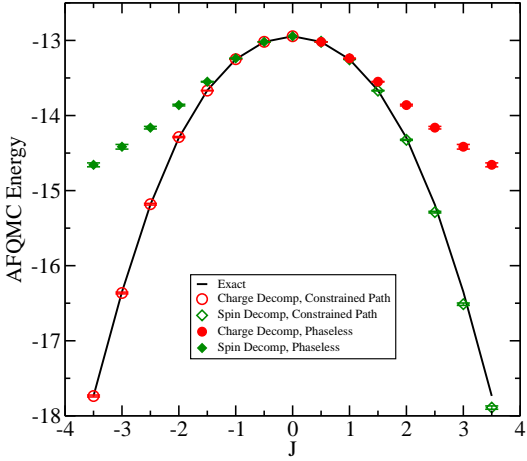}
\captionsetup{font=footnotesize, justification=raggedright, singlelinecheck=false}
\caption{Comparison of phaseless and constrained path AFQMC energy errors as a function of $J$ for a two-band, 5$\times$1 HK model. Open circles denote parameters at which the constrained path approximation was employed, while closed circles denote parameters at which the phaseless approximation was employed. Here, we set N$_{\uparrow}$= N$_{\downarrow}$=6, $t=1$, $U=0$, $U_{1}=0$, and $U_{2}=0$. FE trial wave functions were used for both the initial and trial wave functions, and 560 walkers were employed in each calculation.}
\label{f3}
\end{figure}
%\end{center}
%%%% end figure---------------------------------------
The accuracy of AFQMC predictions are also influenced by the constrained path and phaseless approximations employed. In Figure \ref{f3}, we compare the errors produced by these approximations. As discussed in Section \ref{HSTransform}, for $J>0$, the spin decomposition will yield real propagators that we constrain using the constrained path approximation, while for $J<0$, the spin decomposition will yield complex propagators that we constrain using the phaseless approximation. The charge decomposition behaves in the opposite fashion with respect to $J$. As shown in Figure \ref{f3}, the constrained path approximation behaves significantly better than the phaseless approximation, which appreciably differs from the exact results for $|J|>1.5$. Indeed, the constrained path approximation nearly reproduces the exact results for $J<0$, only manifesting a slight deviation for larger positive values of $J$. These results attest to the fact that using the transformations we describe to prevent the phase problem from emerging is key to maintaining AFQMC accuracy. They also underscore that our method is capable of simulating -$J$ values, which have been unattainable in previous QMC simulations. We expect these trends in accuracy to generalize to models with more bands and higher dimensionality. 

\subsection{Application to Three-Band Hubbard Kanamori Models}
\label{three_band}

\begin{center}
\begin{figure}[htbp]
\includegraphics[width=7.5cm]{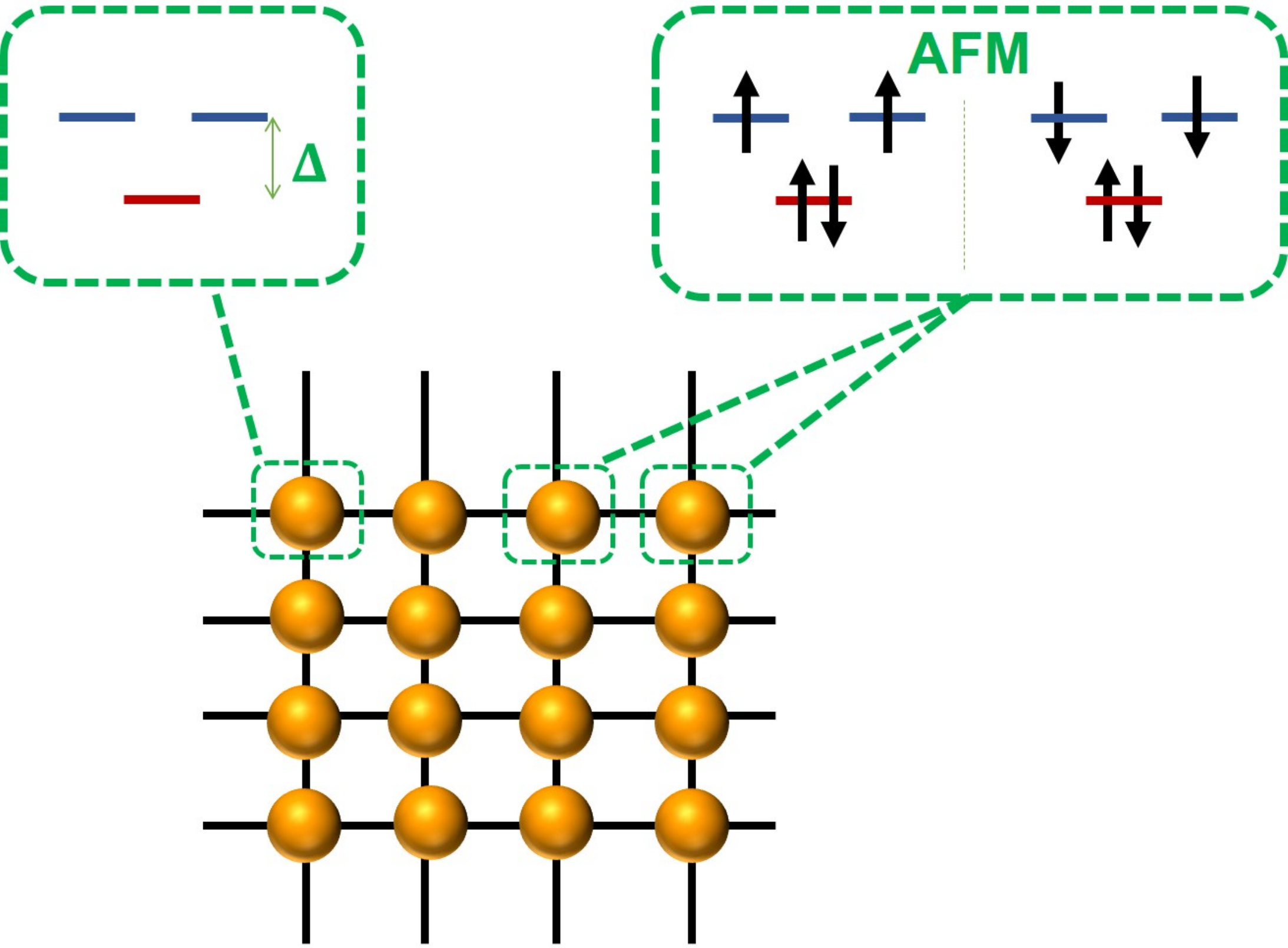}
\captionsetup{font=footnotesize, justification=raggedright, singlelinecheck=false}
\caption{Schematic of our three-band model on a 4x4 lattice. At each site, there is one atom with three bands, one of which is lower in energy by $\Delta$ than the other two degenerate bands. The top right box illustrates a situation in which AFM order is present between adjacent lattice sites.}
\label{three_band_structure}
\end{figure}
\end{center}

In order to understand how our techniques generalize to models that approximate more realistic materials and their magnetic phase transitions, we constructed a three-band model with an adjustable band gap. As illustrated in Figure \ref{three_band_structure}, in this model, three bands are located at each site, one band of which is lower in energy by a `band gap' parameter, $\Delta$, than the other two degenerate bands. When $\Delta = 0$, all three bands are completely degenerate. Similar to the two-band model, the hopping occurs between adjacent sites within the same bands, with hopping constant $t_{ij}=1$. While the band gap would be fixed in any given material, creating a separate $\Delta$ parameter enables us to sample a range of band gaps and, by extension, to drive magnetic ordering transitions. We moreover assume that $U^{i}=U$ and $J^{ij}=J=0.15 U$ with $U_1^{ij}=U_1=U-2J$ and $U_2^{ij}=U_2=U-3J$, which are appropriate for the description of transition-metal oxides with a partially occupied $t_{2g}$ shell.\cite{sugano_tanabe_kamimura_1970} In the following discussion, we fix our filling such that an average of four electrons occupy the three bands at each lattice site.  

%%%% begin figure----------------------------------------------
%%%% end figure---------------------------------------

%%%% begin figure----------------------------------------------
\begin{center}
\begin{figure}[htbp]
\includegraphics[width=10.5cm]{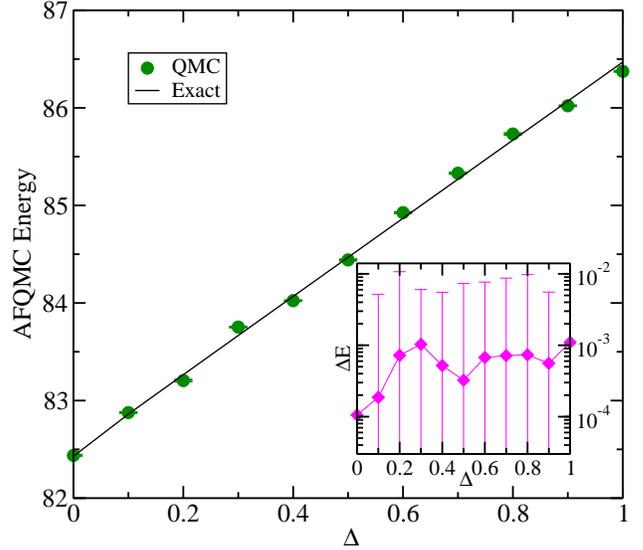}
\captionsetup{font=footnotesize, justification=raggedright, singlelinecheck=false}
\caption{AFQMC ground state energy as a function of the band gap magnitude, $\Delta$, for the three-band, 2$\times$2 HK model using the charge decomposition and GHF trial wave functions. Here, all of the other Hamiltonian parameters are held fixed at $t=1$, $U=6$, $U_1=U-2J$, $U_2=U-3J$, and $J=0.15U$ with $N_{\uparrow}=N_{\downarrow}=8$. Relative errors, $\Delta E$, taken with respect to ED results are plotted in the inset for clarity.}
\label{three_band_mu}
\end{figure}
\end{center}
%%%% end figure---------------------------------------

As an initial step, we benchmarked our AFQMC method against ED results. Diverging from our previous two-band analysis, as part of our three-band benchmarks, we studied our model on two-dimensional lattices with periodic boundary conditions, only varying $\Delta$ and $U$ while keeping the other parameter relationships fixed in order to preserve realism. Our simulations were initialized with 560 walkers and GHF initial and trial wave functions for all of the benchmarks described below. The charge decomposition with the phaseless approximation was employed throughout this section.
%may want to add better explanation for this choice here, if we have one; just sticks out because we said the spin decomposition was better in the previous section

In Figure \ref{three_band_mu}, we illustrate how the energy and relative errors change as $\Delta$ is varied from 0 to 1 with $U=6$ on a 2$\times$2 lattice. At fixed $U$, the relative error remains fairly stable and less than 0.1\% throughout this range. This may be anticipated since the band gap only modifies the magnitude of the one-body terms and does not change the phase of the model, which do not directly contribute to our method's stochastic errors.   
%%%% begin figure----------------------------------------------
\begin{center}
\begin{figure}[htbp]
\includegraphics[width=10.5cm]{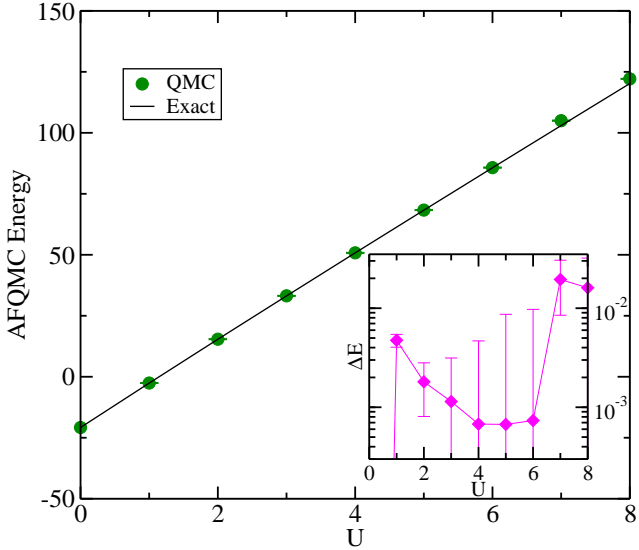}
\captionsetup{font=footnotesize, justification=raggedright, singlelinecheck=false}
\caption{AFQMC ground state energy vs. $U$ for the three-band, 2$\times$2 HK model using the charge decomposition and GHF trial wave functions. Here, all of the other Hamiltonian parameters are held fixed at $t=1$, $\Delta =0.8$, $U_1=U-2J$, $U_2=U-3J$, and $J=0.15U$ with $N_{\uparrow}=N_{\downarrow}=8$. Relative errors, $\Delta E$, are taken with respect to ED results are plotted in the inset for clarity.}
\label{three_band_U}
\end{figure}
\end{center}
%%%% end figure---------------------------------------

In Figure \ref{three_band_U}, instead of scanning $\Delta$, we scan $U$ with $\Delta = 0.8$. As shown in Figure \ref{three_band_U}, the relative errors are larger in this case, but still range from 0.1\% for $U<6$ to 1\% for $U>6$. Errors would be expected to grow in this manner as the system becomes more correlated. Overall, the magnitudes of these relative errors suggest that AFQMC's performance is promising. 

The rationale for introducing the band gap $\Delta$ parameter is to enable tuning of the magnetic order of the model system. Intuitively, when the band gap is small, the three bands are nearly degenerate and the four electrons have the largest freedom to move among the bands. Such a situation would favor ferromagnetic (FM) order. However, when the band gap becomes sufficiently large, two electrons will populate the lower band, forcing the other two electrons to reside on the higher energy bands. Such a situation would favor AFM order. 

\begin{center}
\begin{figure}[htbp]
\includegraphics[width=10.5cm]{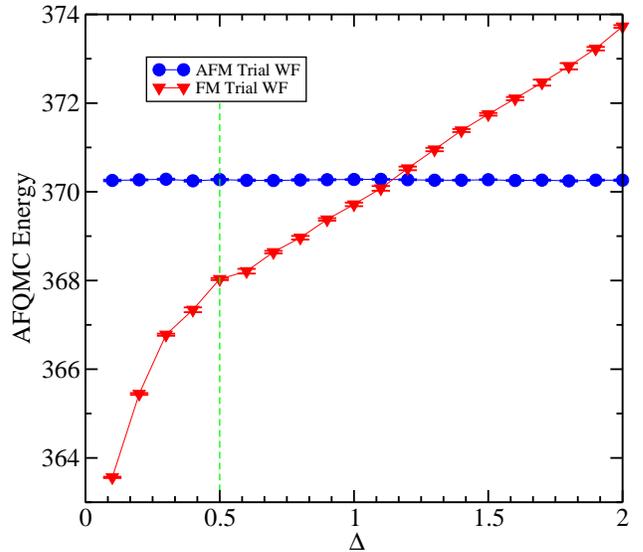}
\captionsetup{font=footnotesize, justification=raggedright, singlelinecheck=false}
\caption{AFQMC ground state energy vs. band gap magnitude, $\Delta$, for the three-band, 4$\times$4 HK model using the charge decomposition. GHF trial wave functions with both FM order and AFM order are used. A QMC predicted phase transition occurred at around $\Delta=1.1$. Hartree-Fock predicted transition point is at $\Delta = 0.5$ illustrated in green dotted line. Here, all of the other Hamiltonian parameters are held fixed at $t=1$, $U=6$, $U_1=U-2J$, $U_2=U-3J$, and $J=0.15$ $U$ with $N_{\uparrow}=N_{\downarrow}=32$.}
\label{three_band_phase}
\end{figure}
\end{center}

This intuition was confirmed by comparing the AFQMC energies attained using trial wave functions with FM and AFM order, respectively (see Figure~\ref{three_band_phase}). Typically, GHF calculations converge to the lowest state with the same magnetic order as the initial state. Thus, in order to construct wave functions with FM order, a randomly initialized density matrix was supplied to the GHF self-consistent equations; to construct wave functions with AFM order, an AFM-ordered initial density matrix was supplied. Several independent GHF calculations were conducted for each system studied to guarantee that the final GHF wave functions produced attained their global minima. At large $\Delta$ ($\Delta \gtrsim 1.1$) values at which ferromagnetic order is disfavored, GHF calculations initialized with random density matrices often developed order. In these situations, FM wave functions produced at smaller values of $\Delta$ were used as trial wave functions in ``FM'' AFQMC calculations performed at larger $\Delta$ values. Figure \ref{three_band_phase} depicts the energies of AFQMC simulations performed with AFM and FM trial wave functions, respectively, as a function of band gap. All of the AFQMC energies presented here are the lowest energies we can obtain at each $\Delta$. At smaller $\Delta$s, trial wave functions with FM order led to the lowest AFQMC energies, while at larger $\Delta$s, AFM trial wave functions did so. This confirms that our model undergoes a ferromagnetic to antiferromagnetic transition at roughly $\Delta = 1.1$. In contrast, Hartree-Fock theory predicts the transition at $\Delta = 0.5$, which is reasonable since Hartree-Fock theory tends to fall into AFM order sector. An illustration of the AFM order exhibited by our model is depicted in Figure~\ref{three_band_structure}.  

%%%% begin figure----------------------------------------------
%%%% end figure---------------------------------------

To further corroborate the phase transition we observe, we extrapolated the magnitude of the charge gap at $\Delta=0.2$ and $\Delta = 1.5$. To do so, we computed the ground state AFQMC energies of 4$\times$4, 6$\times$6, and 8$\times$8-site systems, with three bands filled with four electrons situated at each site. The charge gap may be determined by computing $E_{N-1}+E_{N+1}-2E_{N}$, where $N$ denotes the total number of electrons in the system. To determine the charge gap in the thermodynamic limit, we fit a $1/L$ form, where $L$ denotes the total number of lattice sites, to the energies and extrapolated to the infinite $L$ limit (see the Supplemental Materials for more details). The energies produced using FM initial trial wave functions were used to ascertain the $\Delta = 0.2$ charge gap, while those produced using AFM wave functions were employed to ascertain the $\Delta = 1.5$ charge gap. The charge gaps obtained are presented in Table \ref{charge_gap}. After extrapolations, the $\Delta=0.2$ charge gap converged to -0.006(47) and the $\Delta=1.5$ charge gap converged to 1.201(41). As one would expect antiferromagnetic, not ferromagnetic, order to be accompanied by a charge gap, these extrapolations support our previous conclusions.

%%%% begin table---------------------------------------
\textbf{\begin{table}[htbp]
%\sisetup{separate-uncertainty=true}
%\captionsetup{width=0.5\textwidth}
\footnotesize
\captionsetup{font=footnotesize, justification=raggedright, singlelinecheck=false}
\caption{The charge gaps of the three-band model at $\Delta = 0.2$ and $\Delta=1.5$ for different system sizes calculated using AFQMC. GHF trial wave functions with FM order and AFM order are used at $\Delta = 0.2$ and $\Delta = 1.5$, respectively. All of the other Hamiltonian parameters are held fixed at $t=1$, $U=6$, $U_1=U-2J$, $U_2=U-3J$, and $J=0.15U$. The electron density per band is $4/3$.} 
\label{charge_gap}
\begin{ruledtabular}
\begin{tabular}{ccc}
\# of bands &  Charge Gap ($\Delta = 0.2 $) &  Charge Gap ($\Delta = 1.5 $) \\ 
\hline
    4x4x3  & 0.222(29)  & 1.311(32) \\
    6x6x3 & 0.103(27)  & 1.268(35) \\
    8x8x3  & 0.015(72)  & 1.225(36) \\
    $\infty$ & -0.006(47) & 1.201(41) \\
\end{tabular}
\end{ruledtabular}
\end{table}}
%%%% end table------------------------------------------------

The successful determination of the magnetic order and charge gaps in this model system illustrate our method's promise for accurately modeling realistic materials. 

\section{Conclusions}
\label{conclusions}
In summary, we have presented a ground state AFQMC framework suited for the study of the HK model, a multi-band model designed to capture the Hund's physics of many $d$- and $f$-electron materials. Diverging with past QMC studies of the HK model, we employ a novel set of HS transformations to decouple the Hund's coupling term while preserving the term's essential physics. We find that by carefully combining these transformations with a form of importance sampling that shifts our propagators, well-optimized GHF wave functions, and the constrained path and phaseless approximations, we can accurately predict the energetics of benchmark lattice models and the magnetic order of much larger models that approximate realistic materials. Overall, we find that the phaseless version of our method produces nearly exact energies for small models for $-3<J<3$, a range of $J$ values which contains those commonly observed in experiment. This bodes well for the generalization of our method to other systems. 

Our method may readily be extended to include spin-orbit coupling effects and negative $J$ values, which opens the doors to the highly accurate study of exotic, -$J$ fulleride physics.\cite{Nomura_JPCM_2016,Nomura_Science_2015} In order to describe superconducting physics, our method can be adapted to use superconducting trial wave function forms, including Bardeen-Cooper-Schrieffer\cite{Shi_Zhang_PRA_2015,Carlson_Zhang_PRA_2011} and Hartree-Fock-Bogoliubov\cite{Shi_Zhang_PRB_2017} wave functions. We foresee our method having the most immediate impact as a way to delineate low-temperature phase diagrams currently beyond the reach of DMFT methods.\cite{} As the same transformations and importance sampling techniques may readily be adapted into finite temperature AFQMC formalisms,\cite{Hirsch_PRL_1986,Liu_JCTC_2018,Zhang_PRL_1999} the same methods may be used to develop lower scaling, sign and phase problem free impurity solvers. We look forward to employing our methods to more accurately elucidate the complex many body physics of 4$d$ transition metal oxides such as the ruthenates, rhodates, and molybedenates in the near future. 

\begin{acknowledgments}
H.H., B.R., and H.S. thank Andrew Millis, Antoine Georges, and Shiwei Zhang for stimulating discussions, and Qiang Han for providing data and insight. H.H. and B.R. acknowledge support from NSF grant DMR-1726213 and DOE grant DE-SC0019441. H.S. thanks the Flatiron Institute for research support. The Flatiron Institute is a division of the Simons Foundation. This work was conducted using computational
resources and services at the Brown University Center for Computation and Visualization, the Flatiron Institute, and the Extreme Science and Engineering Discovery Environment (XSEDE).
\end{acknowledgments}

\bibliography{references}

\end{document}